# Third Order Intermodulation Power Estimation for N Sinusoidal Channels

Chit-Sang Tsang
California State University
Long Beach, CA 90840
(562)985-1517
ctsang@csulb.edu

*Abstract* — In this paper analysis is given to find the third order intermodulation power given $N$ sinusoids are fed into a nonlinear device. A simple expression of the third order intermodulation power is given for the case that the center frequencies of the $N$ input sinusoids are equally spaced. Further, if the powers of the $N$ signals are equal, the expression becomes a closed form expression. The analysis will be helpful for communication system engineering in estimating the adjacent channel interference due to nonlinearity. Numerical results are presented for various values of $N$ (number of input channels). Though the analysis assumes the input signals to be sinusoids without phase modulation, the third order intermodulation power estimate serves as a good estimate for link budget computation purpose. For the case that the center frequencies of the input sinusoids are not spaced equally, the analysis can still highly likely be applied if we insert pseudo channels in between the real channels so that all (real and pseudo) channels are spaced equally (or approximately equally for approximation). In this case, the pseudo channel powers are set to zero so that the interference powers due to the pseudo channels will not be included in the analysis. In other words, the analysis is highly likely applicable without the constraint of the input channel center frequencies being equally likely. Simulations are also provided for the case that the input sinusoids are QPSK modulated.

## 1. INTRODUCTION

Nonlinearity effect of signals going through a nonlinear device is always of interest to communication systems especially for the case that multiple channels are fed through the same nonlinear device such as high power amplifiers. For most practical nonlinear amplifiers, only the third order intermodulations are of interest as the higher order nonlinear components are usually negligible. Even number order intermodulations are not of interest as they do not introduce adjacent channel interferences (ACI) to the channels of interest. As a result, system engineers usually focus at the effect of third order intermodulation.

ACI is not a serious problem if there is only one signal feeding into a nonlinear device as undesired frequency components generated after nonlinearity can be kept under control by techniques such as filtering. For communication system engineers special interests are given to passing multiple signals into the same nonlinearity such as high power amplifiers in order to save transmitter hardware cost. For current technology and applications such as satellite communication and cellular telephone systems, it is very common to transmit composite signals through the same power amplifier and transmitter. In that case, the ACI cannot be neglected.

Communication system engineers are interested in the analysis of the ACI powers. With the interference powers over the adjacent channels available, system engineers can perform simple link budget analysis for the communication systems and find out whether any weak channels could be particularly vulnerable to the ACI generated.

Analyses of nonlinear effects on signals through nonlinear devices have been widely studied. The analyses have various levels of complexity for various purposes. Ziemer [1] gave an introduction of the third order intermodulation. Shimbo [2] gave a very detailed mathematical analysis of nonlinearity study with full coverage of AM/AM and AM/PM effects. Sklar [3] gave a treatment from a system engineering view. Jeruchim [4] gave different models of nonlinearity and how they can be simulated.

In this paper multiple signals in the form of $N$ sinusoids are fed into a nonlinear device. Its outputs, including the third order intermodulation products, over the channels are given. For simplicity of analysis, it is assumed that the frequencies of the signals are equally spaced. However, as illustrated below that this assumption does not constitute a serious constraint and can be used for more general cases.

Simulations are also provided for the case that the input sinusoids are Quadrature Phase-Shift Keying (QPSK) signals. The third order intermodulation powers simulated are compared to the analysis assuming no modulation. It is found that the analysis serves as good estimation for the case with QPSK modulation.

## 2. THIRD ORDER INTERMODULATION ANALYSIS OF N CHANNELS

Consider a composite input signal $x(t)$ to be fed into a nonlinear device.

$$x(t) = \sum_{k=1}^{N} A_k \cos(2\pi f_k t + \theta_k) \tag{1}$$

The input signal has $N$ input channels and each has the amplitude $A_k$, frequency $f_k$, and phase $\theta_k$ respectively. The output of the nonlinear device is given by

$$y(t) = \sum_{n=0}^{\infty} \rho_n x^n(t) \tag{2}$$

where $\rho_n$ is the $n^{th}$ coefficient of the nonlinear device. As only the third order intermodulation term is of interest, we only look into the term

$$y_3(t) = \rho_3 x^3(t) \tag{3}$$

It can be shown that

$$\left(\sum_{k=1}^{N} a_k\right)^3 = \sum_{k=1}^{N} a_k^3 + 3\sum_{k=1}^{N}\sum_{\substack{i=1 \\ i\neq k}}^{N} a_k a_i^2 + \sum_{k=1}^{N}\sum_{\substack{i=1 \\ i\neq k}}^{N}\sum_{\substack{j=1 \\ j\neq i,k}}^{N} a_k a_i a_j \tag{4}$$

Let $a_k = A_k \cos(2\pi f_k t + \theta_k)$, and substitute into eq. (4) to get

$$\frac{y_3(t)}{\rho_3} = \sum_{k=1}^{N} A_k^3 \cos^3(2\pi f_k t + \theta_k)$$
$$+ 3\sum_{k=1}^{N}\sum_{\substack{i=1 \\ i\neq k}}^{N} A_k A_i^2 \cos(2\pi f_k t + \theta_k)\cos^2(2\pi f_i t + \theta_i)$$
$$+ \sum_{k=1}^{N}\sum_{\substack{i=1 \\ i\neq k}}^{N}\sum_{\substack{j=1 \\ j\neq i,k}}^{N} A_k A_i A_j \cos(2\pi f_k t + \theta_k)$$
$$\cos(2\pi f_i t + \theta_i)\cos(2\pi f_j t + \theta_j) \tag{5}$$

Applying the trigonometric identities for the products of cosines, eq. (5) can be simplified to

$$\frac{y_3(t)}{\rho_3} = \frac{1}{4}\sum_{k=1}^{N} A_k^3 [3\cos(2\pi f_k t + \theta_k)$$
$$+ \cos(2\pi(3f_k)t + 3\theta_k)]$$
$$+ \frac{3}{2}\sum_{k=1}^{N}\sum_{\substack{i=1 \\ i\neq k}}^{N} A_k A_i^2 [\cos(2\pi f_k t + \theta_k)$$
$$+ \frac{1}{2}\cos(2\pi(2f_i + f_k)t + 2\theta_i + \theta_k)$$
$$+ \frac{1}{2}\cos(2\pi(2f_i - f_k)t + 2\theta_i - \theta_k)]$$
$$+ \frac{1}{4}\sum_{k=1}^{N}\sum_{\substack{i=1 \\ i\neq k}}^{N}\sum_{\substack{j=1 \\ j\neq i,k}}^{N} A_k A_i A_j$$
$$[\cos(2\pi(f_k + f_i + f_j)t + \theta_k + \theta_i + \theta_j)$$
$$+ \cos(2\pi(f_k + f_i - f_j)t + \theta_k + \theta_i - \theta_j)$$
$$+ \cos(2\pi(f_k - f_i + f_j)t + \theta_k - \theta_i + \theta_j)$$
$$+ \cos(2\pi(f_k - f_i - f_j)t + \theta_k - \theta_i - \theta_j) \tag{6}$$

Ignore the frequency terms which are out of the frequency bands of interest, we get

$$\frac{y_3(t)}{\rho_3} = \frac{3}{4}\sum_{k=1}^{N} A_k^3 \cos(2\pi f_k t + \theta_k)$$
$$+ \frac{3}{2}\sum_{k=1}^{N}\sum_{\substack{i=1 \\ i\neq k}}^{N} A_k A_i^2 [\cos(2\pi f_k t + \theta_k)$$
$$+ \frac{1}{2}\cos(2\pi(2f_i - f_k)t + 2\theta_i - \theta_k)]$$
$$+ \frac{1}{4}\sum_{k=1}^{N}\sum_{\substack{i=1 \\ i\neq k}}^{N}\sum_{\substack{j=1 \\ j\neq i,k}}^{N} A_k A_i A_j$$
$$[\cos(2\pi(f_k + f_i - f_j)t + \theta_k + \theta_i - \theta_j)$$
$$+ \cos(2\pi(f_k - f_i + f_j)t + \theta_k - \theta_i + \theta_j)$$
$$+ \cos(2\pi(f_k - f_i - f_j)t + \theta_k - \theta_i - \theta_j)] \tag{7}$$

Define $\quad P_T = \sum_{i=1}^{N} A_i^2 \tag{8}$

Then $\quad \sum_{\substack{i=1 \\ i\neq k}}^{N} A_i^2 = P_T - A_k^2 \tag{9}$

The first term in the second sum of eq. (7) can be written as

$$\frac{3}{2}\sum_{k=1}^{N}\sum_{\substack{i=1 \\ i\neq k}}^{N} A_k A_i^2 \cos(2\pi f_k t + \theta_k)$$
$$= \frac{3}{2}\sum_{k=1}^{N} A_k (P_T - A_k^2) \cos(2\pi f_k t + \theta_k) \tag{10}$$

Combining the term in eq. (10) with the first term in eq. (7) to give

$$\frac{y_3(t)}{\rho_3} = \sum_{k=1}^{N} (\frac{3}{4} A_k^3 + \frac{3}{2} A_k (P_T - A_k^2))\cos(2\pi f_k t + \theta_k)$$
$$+ \frac{3}{4}\sum_{k=1}^{N}\sum_{\substack{i=1 \\ i\neq k}}^{N} A_k A_i^2 \cos(2\pi(2f_i - f_k)t + 2\theta_i - \theta_k)$$
$$+ \frac{1}{4}\sum_{k=1}^{N}\sum_{\substack{i=1 \\ i\neq k}}^{N}\sum_{\substack{j=1 \\ j\neq i,k}}^{N} A_k A_i A_j$$
$$[\cos(2\pi(f_k + f_i - f_j)t + \theta_k + \theta_i - \theta_j)$$
$$+ \cos(2\pi(f_k - f_i + f_j)t + \theta_k - \theta_i + \theta_j)$$
$$+ \cos(2\pi(f_k - f_i - f_j)t + \theta_k - \theta_i - \theta_j)] \tag{11}$$

The first term in eq. (11) is the same as the desirable signals and the rest are ACI terms caused by the third order intermodulation. Denote the ACI terms by $I_3$.

$$I_3 = \frac{3}{4}\sum_{k=1}^{N}\sum_{\substack{i=1 \\ i\neq k}}^{N} A_k A_i^2 \cos(2\pi(2f_i - f_k)t + 2\theta_i - \theta_k)$$
$$+ \frac{1}{4}\sum_{k=1}^{N}\sum_{\substack{i=1 \\ i\neq k}}^{N}\sum_{\substack{j=1 \\ j\neq i,k}}^{N} A_k A_i A_j$$
$$[\cos(2\pi(f_k + f_i - f_j)t + \theta_k + \theta_i - \theta_j)$$
$$+ \cos(2\pi(f_k - f_i + f_j)t + \theta_k - \theta_i + \theta_j)$$
$$+ \cos(2\pi(f_k - f_i - f_j)t + \theta_k - \theta_i - \theta_j)] \tag{12}$$

Assume the frequencies of the channels are equally spaced, i.e.,

$$f_k = f_0 + (k-1)\Delta f \quad ; k = 1,..,N \tag{13}$$

where $f_0$ is an arbitrary constant reference frequency, and $\Delta f$ is the frequency separation between two adjacent channels. Given a particular channel $n$, we are interested to find the ACI power resided in the channel.

Denote the double sum term in eq. (12) as $I_{3D}$.

$$I_{3D} = \frac{3}{4}\sum_{k=1}^{N}\sum_{\substack{i=1 \\ i\neq k}}^{N} A_k A_i^2 \cos(2\pi(2f_i - f_k)t + 2\theta_i - \theta_k) \quad (14)$$

The frequencies of the signal in eq. (14) can be written as

$$2f_i - f_k = f_0 + (2i - k - 1)\Delta f \quad (15)$$

Given a particular channel $n$, its frequency is $f_n = f_0 + (n-1)\Delta f$. If the frequencies in eq. (15) are to reside in channel $n$, then the constraint imposed on the channel numbers $k$ and $i$ in eq. (15) is

$$k = 2i - n \quad (16)$$

Given the values of $k$ and $n$, $i = (k+n)/2$. Since $i$ is an integer, $k + n$ must be an even number. Besides, the constraint $i \neq k$ imposed by eq. (14) gives

$$i = (k+n)/2 \neq k \quad \text{or} \quad k \neq n \quad (17)$$

With those conditions set, eq. (14) can be simplified to

$$I_{3D}(n) = \frac{3}{4}\sum_{\substack{k=1, k\neq n \\ k+n=even}}^{N} A_k A_{(k+n)/2}^2 \cos(2\pi f_n t + 2\theta_{(k+n)/2} - \theta_k) \quad (18)$$

We include the channel number $n$ in $I_{3D}(n)$ to indicate the ACI component residing in channel $n$. We next consider the first triple sum term in eq. (12).

$$I_{3T1}(n) = \frac{1}{4}\sum_{k=1}^{N}\sum_{\substack{i=1 \\ i\neq k}}^{N}\sum_{\substack{j=1 \\ j\neq i,k}}^{N} A_k A_i A_j \cos(2\pi(f_k + f_i - f_j)t + \theta_k + \theta_i - \theta_j) \quad (19)$$

With the similar reasoning as deriving eq. (18), eq. (19) can be simplified to

$$I_{3T1}(n) = \frac{1}{4}\sum_{\substack{k=1 \\ k\neq n}}^{N}\sum_{\substack{i=1 \\ i\neq k,n \\ i>n-k \\ i\leq N+n-k}}^{N} A_k A_i A_{k+i-n} \cos(2\pi f_n t + \theta_k + \theta_i - \theta_{k+i-n}) \quad (20)$$

The constraints $i > n - k$ and $i \leq N + n - k$ in the second sum is due to the constraint $1 \leq k + i - n \leq N$, i.e., all channel numbers have to be between 1 and $N$. Similarly the second and third terms in the triple sum in eq. (12) can be simplified to

$$I_{3T2}(n) = \frac{1}{4}\sum_{\substack{k=1 \\ k\neq n}}^{N}\sum_{\substack{i=1 \\ i\neq k,2k-n \\ i>k-n \\ i\leq N+k-n}}^{N} A_k A_i A_{i-k+n} \cos(2\pi f_n t + \theta_k - \theta_i + \theta_{i-k+n}) \quad (21)$$

$$I_{3T3}(n) = \frac{1}{4}\sum_{k=1}^{N}\sum_{\substack{i=1 \\ i\neq k,n,(k+n)/2 \\ i<k+n \\ i\geq k+n-N}}^{N} A_k A_i A_{k-i+n} \cos(2\pi f_n t + \theta_k - \theta_i - \theta_{k-i+n}) \quad (22)$$

With the expressions derived for $I_{3D}(n)$, $I_{3T1}(n)$, $I_{3T2}(n)$, and $I_{3T3}(n)$, we are ready to find the ACI powers resided in a particular channel $n$ of interest. The powers can be computed numerically if $A_k$ are not the same for different $k$. Since the ACI terms are all cosine waveforms with different phases, the total power is simply the sum of the individual powers. For the rare case that all phases are the same, then we need to sum the power coherently, i.e., all cosines have the same frequency ($f_n$) and phases, we first sum together the amplitudes before taking the square. Obviously, this is a very rare case and is not of interest in practice.

## 3. EQUAL CHANNEL POWERS

For the case $A_k$ are identical for all $k$, the expressions of $I_{3D}(n)$, $I_{3T1}(n)$, $I_{3T2}(n)$, and $I_{3T3}(n)$ can further be simplified. Consider eq. (18) with $A_k = A$.

$$I_{3D}(n) = \frac{3}{4}A^3 \sum_{\substack{k=1, k\neq n \\ k+n=even}}^{N} \cos(2\pi f_n t + 2\theta_{(k+n)/2} - \theta_k) \quad (23)$$

Its power, assuming unequal phases $\theta_k$, becomes

$$P_{I3D}(n) = \frac{L_D(n)}{2}\left(\frac{3}{4}A^3\right)^2 \quad (24)$$

where $L_D(n) = \sum_{\substack{k=1, k\neq n \\ k+n=even}}^{N} 1 \quad (25)$

Its value is given in Table 1 for various cases of $N$ and $n$.

Table 1 $L_D(n)$ Values

| $N$ | $n$ | $L_D(n)$ |
|---|---|---|
| even | even or odd | $(N-2)/2$ |
| odd | even | $(N-3)/2$ |
| | odd | $(N-1)/2$ |

Consider eq. (20) with $A_k = A$.

$$I_{3T1}(n) = \frac{1}{4}\sum_{\substack{k=1 \\ k\neq n}}^{N}\sum_{\substack{i=1 \\ i\neq k,n \\ i>n-k \\ i\leq N+n-k}}^{N} A^3 \cos(2\pi f_n t + \theta_k + \theta_i - \theta_{k+i-n}) \quad (26)$$

Its power, assuming unequal phases $\theta_k$, becomes

$$P_{I3T1}(n) = x\frac{L_T(n)}{2}\left(\frac{A^3}{4}\right)^2 \quad (27)$$

where $L_T(n) = \sum_{\substack{k=1 \\ k \neq n}}^{N} \sum_{\substack{i=1 \\ i \neq k,n \\ i > n-k \\ i \leq N+n-k}}^{N} 1$ (28)

and $x = 6$

To evaluate $L_T(n)$ in eq. (28), it is helpful to plot $i$ versus $k$ for $1 \leq i, k \leq N$. In the plot mark the points satisfying the summing conditions indicated in eq. (28). From the geometry of the point patterns in the plot, it is possible to obtain the closed form expression of $L_T(n)$. It is found that

$$L_T(n) = 2 + \frac{1}{2}(N^2 + 2nN - 5N - 2n^2 + 2n) - floor((N+n)/2) + floor(n/2) \quad (29)$$

where the function $floor(y)$ rounds the value of y to the nearest integer less than y.

In eq. (27) the power is evaluated assuming all the cosines in eq. (26) have different phases. In evaluating $L_T(n)$ it is found that the cosines, satisfying the condition of eq. (28), come in groups of 6 cosines with the same frequencies and phases. These 6 cosines are summed together in the form of $6\cos(2\pi f_n t + \theta_k + \theta_i - \theta_{k+i-n})$. Its power is $6^2/2=18$ instead of $6(1^2/2)=3$. The difference is a factor of $18/3=6$. Therefore, we need to include the factor $x = 6$ in eq. (27).

It can also be shown that $I_{3T1}(n)$, $I_{3T2}(n)$, and $I_{3T3}(n)$ have the same number of terms in the double sums, and for the case $A_k = A$ their powers are equal. Therefore, summing together the ACI powers contributed by $I_{3D}(n)$, $I_{3T1}(n)$, $I_{3T2}(n)$, and $I_{3T3}(n)$, we get

$$P_{I3}(n) = P_{I3D}(n) + 3P_{I3T1}(n) = \frac{9}{32}(L_D(n) + 2L_T(n))A^6 \quad (30)$$

## 4. UNEQUAL CHANNEL FREQUENCY SPACING

The above analysis assumes all the channels have equal channel spacing. It might sound restrictive on this assumption. However, with minor modification, the analysis is highly likely applicable to the unequal channel frequency spacing scenario.

The idea is to make up pseudo channels to fill up the frequency gaps in between so that the overall system has equal, or approximately equal, frequency spacing. The pseudo channels have zero amplitudes and thus will not contribute to actual ACI power computation. In that way, the analysis in Section 2 could be applicable.

As an illustration, consider the scenario below.
- 4 channels with frequencies $f_{c1} = 5$, $f_{c2} = 7$, $f_{c3} = 10$, and $f_{c4} = 14$.
- Make up a system of $N = 10$ channels with 6 fictitious channels. The pseudo channels have zero amplitudes (zero power).
- New system consists of: $f_1 = 5$, $f_2 = 6$, $f_3 = 7$, $f_4 = 8$, $f_5 = 9$, $f_6 = 10$, $f_7 = 11$, $f_8 = 12$, $f_9 = 13$, and $f_{10} = 14$.
- The original 4 channels take on $f_1$, $f_3$, $f_6$, and $f_{10}$ respectively.

Apply the analysis in Section 2 and the ACI analysis for this unequal channel frequency spacing system is resolved. Closed form equations in Section 3 assume all channels have equal powers and do not apply in this scenario.

## 5. NUMERICAL RESULTS FOR EQUAL CHANNEL POWERS

In this section numerical results are provided for ACI power assuming all channel powers are equal. For convenience, the normalized ACI power, defined by $P_{I3}/N^2$, is used in the numerical results.

Fig. 1 shows a typical waveform and power spectrum for $x^3(t)$ for $x(t)$ being a sum of three sinusoids. The top two figures show the waveform and spectrum of $x(t)$. The middle figures show the waveform and spectrum of $x^3(t)$. The bottom two figures show the waveform and spectrum of the intermodulation components.

Figs. 2 to 5 plot the normalized ACI power versus the channel number $n$ for the total channel number $N$ equals to 9, 10, 31, and 99 respectively. The ACI has a maximum value at the center of the channels. It means that there is a disadvantage for a channel located at the center of the channel group as it experiences more ACI power compared to the channels located on the sides. For $N$ even, the plot is symmetric about the mid values $N/2$ and $\frac{N}{2} + 1$. For $N$ odd, the plot is symmetric about the mid value $(N + 1)/2$. As $N$ increases, the plot becomes smooth. The symmetry behavior is as expected as there is no difference for channels residing on either frequency sides.

The maximum normalized ACI power is plotted in Fig. 6 versus the total number of channels $N$. The smallest $N$ value shown is 3 as for $N = 2$ there is no third order ACI generated that resides on any one of the two original channels. For a wide range of $N$ the normalized ACI powers are within 0.43. In other words, maximum ACI due to third order intermodulation is bounded by $0.43N^2$.

The ratio of the maximum and minimum ACI powers, $P_{max}/P_{min}$, is of interest in system engineering. It is plotted in Fig. 7 versus $N$. It is seen that the ratio has a lower bound of 1.5, which occurs for large $N$. The ratio takes the largest value of 4 for $N = 3$ and decreases quickly when $N$ increases.

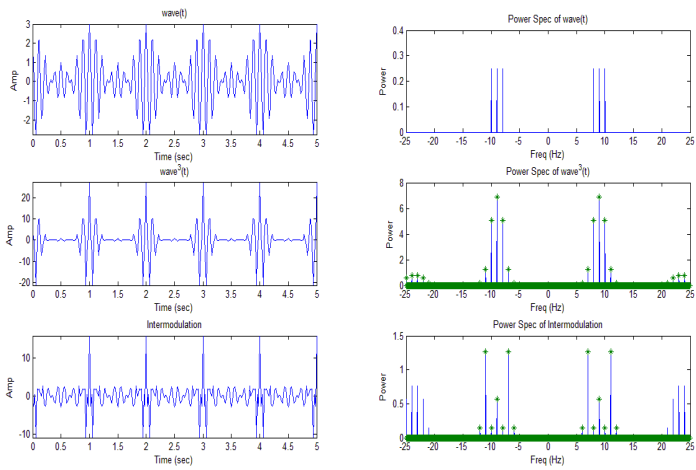

Fig. 1 Intermodulation Waveform and Spectrum

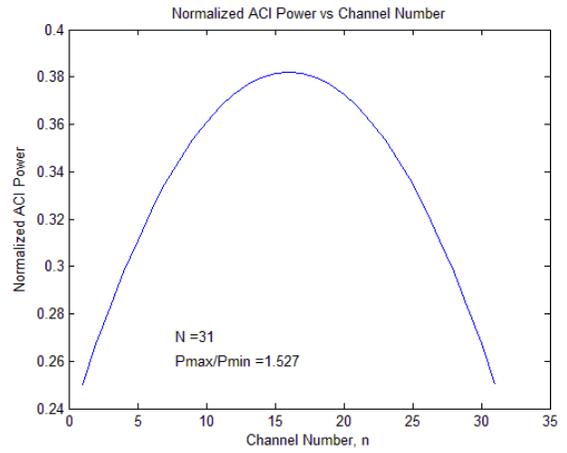

Fig. 4 Normalized ACI Power for $N=31$

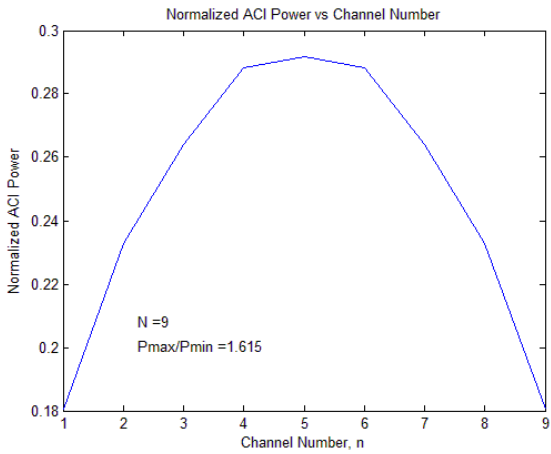

Fig. 2 Normalized ACI Power for $N=9$

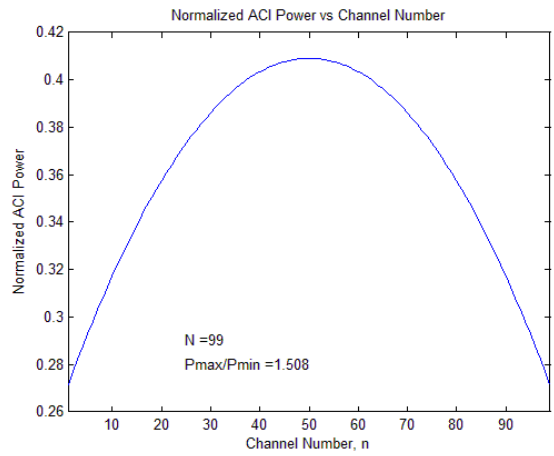

Fig. 5 Normalized ACI Power for $N=99$

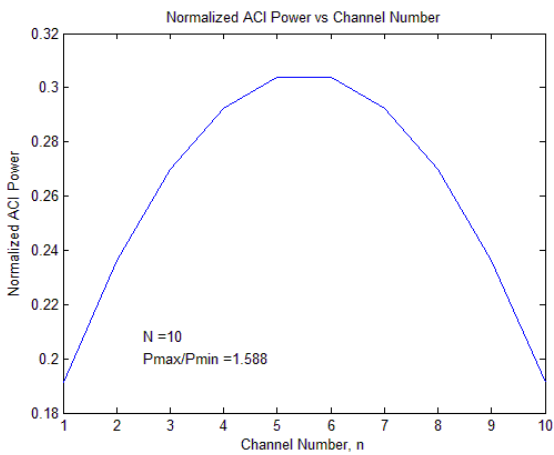

Fig. 3 Normalized ACI Power for $N=10$

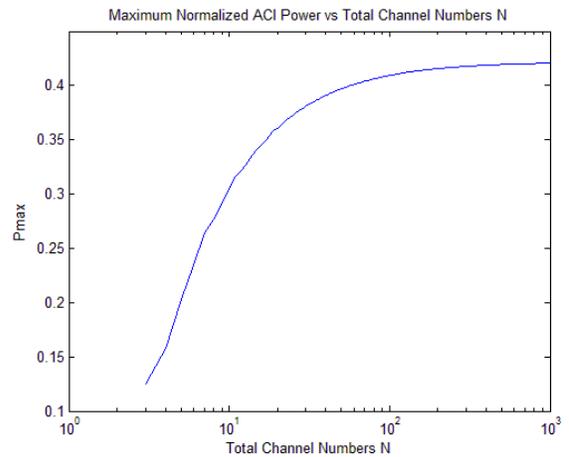

Fig. 6 Maximum Normalized ACI Power

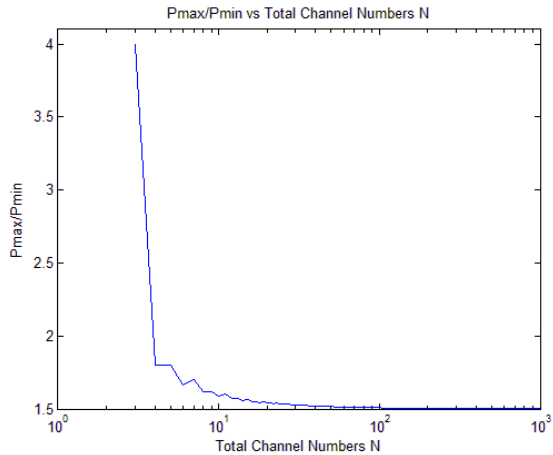

Fig. 7 $P_{max}/P_{min}$ versus $N$

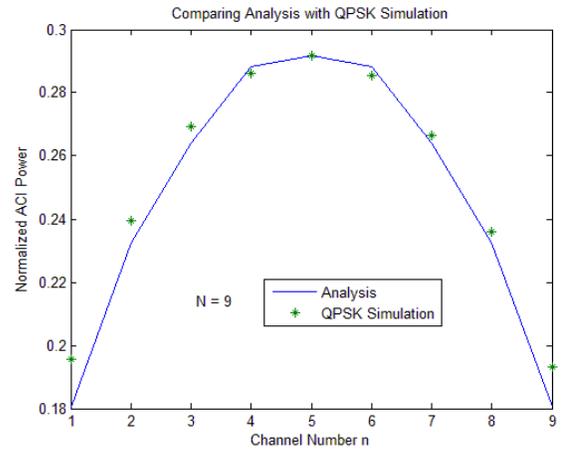

Fig. 9 Comparing ACI Analysis with QPSK Simulation

## 6. COMPARISON WITH QPSK SIMULATION

Simulation was performed to find the third order intermodulation ACI of QPSK signals. Fig. 8 illustrates the power spectrum of 5 composite signals, the cube of the composite signal, and their intermodulation components. Fig. 9 gives a comparison between the analysis and simulation for $N = 9$. For comparison, the simulation result is normalized to match the analysis result at the center ($n = 5$). The other 8 channel ACI results are pretty close to each other.

## 7. SUMMARY AND CONCLUSION

This paper gives an ACI analysis of third order intermodulation of composite sinusoids. Closed form expression is given for the case when all sinusoid amplitudes are identical and same frequency spacing. The analysis is compared with simulation of QPSK signals. With the analysis, system engineers can estimate the effect due to nonlinearity effects on multiple signals passing through a common nonlinear device.

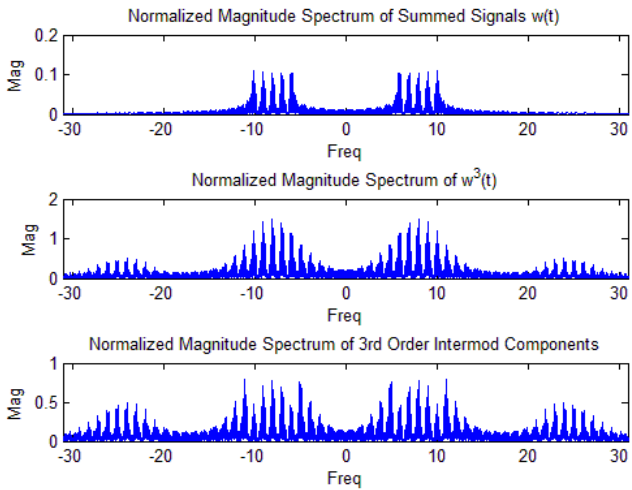

Fig. 8 Intermodulation of 5 QPSK Waveforms and Spectrum

## BIOGRAPHY

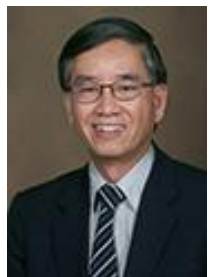

Dr. Chit-Sang Tsang (Senior Member, IEEE) received his Ph.D. in Electrical Engineering from University of Southern California. He was with LinCom Corporation working in NASA space programs. He joined the Electrical Engineering Department of California State University, Long Beach, as Associate Professor in 1988 and is currently Professor of Electrical Engineering. His areas of interest in digital communications are communications system engineering, wireless communications, and synchronization. He also publishes in speech signal processing and neural networks.